\documentclass[superscriptaddress,keywords]{revtex4}
\usepackage{graphicx}
\usepackage{color}

\usepackage{amsmath}
\usepackage{amsfonts}
\usepackage{amssymb}

\newcommand{\be}{\begin{equation}}
\newcommand{\ee}{\end{equation}}

\usepackage{bm} 

\begin{document}

\title{Identification of non-ordinary mesons
from the dispersive connection between their poles and their Regge trajectories:
the $f_0(500)$ resonance} 

\author{ J.T.~Londergan}
\affiliation{Center for Exploration of Energy and Matter, Indiana University, Bloomington, IN 47403, USA}
\affiliation{Physics Department  Indiana University, Bloomington, IN 47405, USA }
\author{J.~Nebreda}
\affiliation{Center for Exploration of Energy and Matter, Indiana University, Bloomington, IN 47403, USA}
\affiliation{Physics Department  Indiana University, Bloomington, IN 47405, USA }
\affiliation{Departamento de F\'isica Te\'orica II, Universidad Complutense de Madrid, 28040 Madrid, Spain}
\affiliation{Yukawa Institute for Theoretical Physics, Kyoto University, 606-8502 Kyoto, Japan
} 
\author{J.R.~Pelaez}
\affiliation{Departamento de F\'isica Te\'orica II, Universidad Complutense de Madrid, 28040 Madrid, Spain}
\author{A.P.~Szczepaniak} 
\affiliation{Center for Exploration of Energy and Matter, Indiana University, Bloomington, IN 47403, USA}
\affiliation{Physics Department  Indiana University, Bloomington, IN 47405, USA }
\affiliation{  Jefferson Laboratory, 12000 Jefferson Avenue, Newport News, VA 23606, USA }
\date{\today}

\begin{abstract}
We show how the Regge trajectory of a resonance can be obtained
from its pole in a scattering process and analytic constraints in the complex angular momentum plane.
The method is suited for resonances that dominate an elastic scattering amplitude.
In particular, from the $\rho(770)$ resonance pole 
in $\pi\pi$ scattering, we obtain its linear Regge trajectory,
characteristic of ordinary quark-antiquark states. 
In contrast, the $f_0(500)$ pole--the sigma meson--which dominates 
scalar isoscalar $\pi\pi$ scattering, yields a
non-linear trajectory with a much smaller slope at the $f_0(500)$ mass. Conversely, 
imposing a linear Regge trajectory for the $f_0(500)$, 
with a slope of typical size, 
yields an elastic amplitude at odds with the data. This provides 
strong support for the non-ordinary nature of the sigma meson.
\end{abstract}

\maketitle

\hspace{1.5cm}{\it Keywords: Regge theory, Light scalar mesons.}

\section{Introduction}
There is growing evidence for the existence of hadrons that fall beyond 
the ordinary quark-antiquark classification of mesons or three quark classification of baryons. 

Most current investigations of dynamical models underlying resonance formation  
 focus on individual partial waves.  This allows to study poles of amplitudes in the complex energy plane at fixed angular momentum and their spectroscopic classification into SU(3) multiplets, which by itself provides limited information about their composition. 
 In this work we take advantage of the analytical properties of amplitudes in the complex angular momentum plane and this enables us to investigate 
  the dynamical linkage of resonances of different spins. The function connecting such 
 resonances is known as the Regge trajectory and its form can be used to discriminate between the underlying (QCD) mechanisms responsible for generating the resonances. 
 For example,  linear $(J,M^2)$
trajectories relating the angular momentum $J$ and the mass squared are naively and intuitively interpreted in terms of the rotation 
of the flux tube connecting a quark and an antiquark. Strong deviations from this linear behavior would suggest a rather different nature.

For illustration, we will apply our method to light resonances in elastic $\pi\pi$ scattering. We consider the $\rho(770)$, which suits well the 
ordinary meson picture, and also 
the $f_0(500)$ or $\sigma$ meson, whose nature, spectroscopic classification, and even its  existence have been the subject of a longstanding debate. 
Apart from its significant role in our understanding
of the spontaneous chiral symmetry breaking of QCD, the nucleon-nucleon attraction, or even the identification of the lightest glueball, our interest in the $f_0(500)$ is that different approaches \cite{Jaffe:1976ig,Pelaez:2006nj} suggest that it
may not be an ordinary quark-antiquark meson. Furthermore, the $\sigma$ meson is often omitted from $(J,M^2)$ trajectory fits \cite{Anisovich:2000kxa}, 
since it does not ``fit well into this classification" or, as in \cite{Masjuan:2012gc}, because it has a large width and it contributes little to the $\chi^2$, but was included in the n-trajectories.

The input for
our approach is  just the position and residue
of the resonance poles in $\pi\pi$ scattering. 
Our $f_0(500)$ choice is then even more pertinent, because
for long its pole parameters 
have been plagued by systematic uncertainties. However, 
recent and rigorous dispersive analyses on scattering data
have provided a model-independent and accurate determination of the $f_0(500)$ \cite{Caprini:2005zr,GarciaMartin:2011jx},
finally settling \cite{PDG} the controversy on its existence.

Even though we shall not be able to compute trajectories over a large energy range, the local behavior can be quite telling. In practice we aim at obtaining the slope and the intercept of the Regge trajectory where the $f_0(500)$ lies, showing the striking differences with 
the $\rho(770)$, thus explaining why it does not fit into the ordinary linear trajectory classification.

This work focuses on properties of  Regge poles of the scattering amplitude.  In this context, our working definition of an 
 ordinary meson is that it correspond to a pole which lies on a Regge trajectory that is almost real and linear with slope of the order of 1$\mbox{GeV}^{-2}$.  We show that Regge trajectory of the  $\rho(770)$  satisfies this criterion, while that of the $f_0(500)$ does not.  Since  trajectories of ordinary mesons are qualitatively understood in terms of conventional quark-antiquark dynamics, 
  the fact that the $f_0(500)$ trajectory is different suggests that a different  mechanism is responsible for the formation 
   of the $f_0(500)$, although at present we cannot specify what that mechanism might be.

\section{Regge trajectory from a single pole}

Let us recall that within analytic S-matrix theory, the method of imposing unitarity constraints from the crossed channels on the direct channel is that of analytic continuation of partial waves into the complex angular momentum plane. Singularities in the angular momentum plane, {\it e.g.} Regge poles, interpolate between direct and crossed-channel dynamics,  containing the most complete description of resonance parameters.

\subsection{Analytic constraints on the trajectory and the residue}

An elastic $\pi\pi$ partial wave near a Regge pole reads
\be
t_l(s)  = \beta(s)/(l-\alpha(s)) + f(l,s),
\label{Reggeliket}
\ee
where $f(l,s)$ is a regular function of $l$, whereas the Regge trajectory $\alpha(s)$ and 
residue $\beta(s)$
satisfy  $\alpha(s^*)=\alpha^*(s)$, $\beta(s^*)=\beta^*(s)$, in the complex-$s$ plane cut along the real axis for $s > 4m_\pi^2$. Note that the pole appears in the second Riemann sheet of
$t_l(s)$, which we normalize as  
\be
t_l(s) =  e^{i\delta_l(s)}\sin{\delta_l(s)}/\rho(s), \quad \rho(s) = \sqrt{1-4m_\pi^2/s} \ ,
\ee
with  $\delta_l(s)$ being the phase shift.  Now, if the pole dominates in Eq.\eqref{Reggeliket}, the unitarity condition above threshold $\mbox{Im}t_l(s)=\rho(s)|t_l(s)|^2$ analytically continued to complex $l$ implies that, for real $l$, 
\be
\mbox{Im}\,\alpha(s)   = \rho(s) \beta(s).   \label{unit} 
\ee
For integer-$l$ partial waves the unitarity relation gives the prescription for how to 
 analytically continue $t_l(s)$ below the elastic cut for $s> 4m_\pi^2$. Similarly,
  Eq.~\eqref{unit}  determines the continuation of $\alpha(s)$, which we will use when studying resonance poles that occur at fixed, integer $l$ and complex $s$.  
  
Note that, if $\beta(s)$ was known, we could use a dispersion relation to determine $\alpha(s)$. Therefore we first discuss the 
analytic properties of the former~\cite{Chu:1969ga}. Near threshold, partial waves behave as $t_l(s) \propto q^{2l}$, where $q^2=s/4-m_\pi^2$ and thus, 
$\beta(s) \propto q^{2\alpha(s)}$.    
Moreover, since the Regge pole contribution to the full amplitude
is proportional to $(2\alpha + 1) P_\alpha(z_s)$, where $z_s$ is the s-channel scattering angle, in order to cancel poles of the Legendre function $P_\alpha(z_s)\propto\Gamma(\alpha + 1/2)$ the residue has to vanish when 
  $\alpha + 3/2$ is a negative integer, i. e.,
\be
\beta(s) =  \gamma(s) \hat s^{\alpha(s)} /\Gamma(\alpha(s) + 3/2) , \label{reduced} 
\ee
where $\hat s =( s-4m_\pi^2)/s_0$. The dimensional scale $s_0=1\,$ GeV$^2$ is introduced for 
convenience and the reduced residue $\gamma(s)$ is, once again, a real 
analytic function. Since on the real axis $\beta(s)$ is real, the phase of $\gamma$ is
\be
\mbox{arg}\,\gamma(s) = - \mbox{Im}\,\alpha(s) \log(\hat s) + \arg \Gamma(\alpha(s) + 3/2). 
\ee
Analyticity therefore demands that 
\be
\gamma(s) = P(s) \exp\left(c_0 + c' s + \frac{s}{\pi} \int_{4m_\pi^2}^\infty \!\!\!\!ds' \frac{\mbox{arg}\,\gamma(s')}{s' (s' - s)} \right), \label{g}
\ee
where $P(s)$ is an entire function. 
Given that we use the elastic approximation, the behavior at large $s$ cannot be determined from first principles. However as we expect linear Regge trajectories to emerge for the $\rho(770)$, we should allow $\alpha$ to behave as a first order polynomial at large-$s$. This  implies that $\mbox{Im}\,\alpha(s)$ decreases with growing $s$ 
and thus $\alpha(s)$ obeys the dispersion relation~\cite{Collins-PLB}, 
\be
\alpha(s) = \alpha_0 + \alpha' s + \frac{s}{\pi} \int_{4m_\pi^2}^\infty ds' \frac{ \mbox{Im}\,\alpha(s')}{s' (s' -s)}. \label{alphadisp}
\ee
   To match the asymptotic behavior of $\beta(s)$ and $\mbox{Im}\,\alpha(s)$   (assuming $\alpha' \ne 0$), it 
 follows from the unitarity equation, Eq.\eqref{unit},  that 
$c' = \alpha' ( \log(\alpha'  s_0) - 1)$ and that $P(s)$ can at most be a constant.    
  Hence, together with 
Eq.~\eqref{Reggeliket}, the following three equations define the ``constrained Regge-pole'' amplitude~\cite{Chu:1969ga}:
\begin{align}
\mbox{Re}\, \alpha(s) & =   \alpha_0 + \alpha' s +  \frac{s}{\pi} PV \int_{4m_\pi^2}^\infty ds' \frac{ \mbox{Im}\,\alpha(s')}{s' (s' -s)}, \label{iteration1}\\
\mbox{Im}\,\alpha(s)&=  \frac{ \rho(s)  b_0 \hat s^{\alpha_0 + \alpha' s} }{|\Gamma(\alpha(s) + \frac{3}{2})|}
 \exp\Bigg( - \alpha' s[1-\log(\alpha' s_0)]
+ \frac{s}{\pi} PV\!\int_{4m_\pi^2}^\infty\!\!ds' \frac{ \mbox{Im}\,\alpha(s') \log\frac{\hat s}{\hat s'} + \mbox{arg }\Gamma\left(\alpha(s')+\frac{3}{2}\right)}{s' (s' - s)} \Bigg), 
\label{iteration2}\\
 \beta(s) &=    \frac{ b_0\hat s^{\alpha_0 + \alpha' s}}{\Gamma(\alpha(s) + \frac{3}{2})} 
 \exp\Bigg( -\alpha' s[1-\log(\alpha' s_0)] 
+  \frac{s}{\pi} \int_{4m_\pi^2}^\infty \!\!ds' \frac{  \mbox{Im}\,\alpha(s') \log\frac{\hat s}{\hat s'}  + \mbox{arg }\Gamma\left(\alpha(s')+\frac{3}{2}\right)}{s' (s' - s)} \Bigg),
 \label{betafromalpha}
 \end{align}
where $PV$ denotes ``principal value''.  Note that Eq.\eqref{betafromalpha} reduces to Eq.\eqref{unit} for real $s$. 

For the $\sigma$-meson, $\beta(s)$ at low energies should also
include the Adler-zero required by chiral symmetry. 
In practice it is enough to place it
at the leading order chiral perturbation theory result \cite{chpt}, i.e.,
 $\beta(s) \propto 2s - m_\pi^2$.
This should be done without spoiling the large $s$-behavior, 
which can be achieved by replacing $\Gamma(\alpha + 3/2)$ by $\Gamma(\alpha + 5/2)$. Such modification leaves the pole at $\alpha(s) = - 3/2$ uncanceled. This is not an issue
since for the $\sigma$ trajectory, $\alpha(s) = \alpha_\sigma(s)$ this pole will 
be located far outside the range of applicability of 
our approach. Hence, for the $f_0(500)$ we should just multiply
the right hand side of Eq.\eqref{iteration2} by $2s-m_\pi^2$ and replace the $3/2$ by $5/2$ inside the gamma functions. Note that $b_0$ now is not dimensionless.

\subsection{Numerical Analysis}
  
We solve the equations for $\alpha(s)$ and $\beta(s)$ numerically. The only inputs are the pole positions $s_M$ and residues $|g_M|$ for the $M= \rho(770)$ and the  $f_0(500)$ resonances.  
 Specifically, the poles are used as input for determining the $\alpha_0, \alpha',b_0$ parameters of the corresponding Regge trajectories,  by requiring that at the pole, on the second Riemann sheet,
$\beta_M(s)/(l  - \alpha_M(s))\rightarrow |g^2_M|/(s-s_M)$, 
with $l=0,1$ for $M=\sigma,\rho$.  We minimize 
the sum of the squared differences between the input and output values 
for the real and imaginary parts of the pole position and for the absolute value of the squared coupling, divided by the corresponding squared uncertainties. The pole parameters are taken from a precise dispersive representation of $\pi\pi$ scattering data \cite{GarciaMartin:2011jx,GarciaMartin:2011cn} that enables a model independent, analytic continuation of partial wave amplitudes to the complex energy plane.   

 For a given set of $\alpha_0, \alpha'$ and $b_0$ parameters
we solve the system of Eqs.~\eqref{iteration1} and \eqref{iteration2},
with the modification due to the Adler-zero in the scalar case, as discussed above. 
This is done by setting $\mbox{Im}\,\alpha( s)=0$ initially, which yields Re$\,\alpha(s)$ using Eq.~\eqref{iteration1}. Then, these real and imaginary parts of $\alpha$ are used in Eq.~\eqref{iteration2} to obtain $\mbox{Im}\,\alpha(s)$. This process is iterated until 
the results converge. Thus, we obtain a constrained Regge-pole amplitude, under the approximation that it is dominated by a single Regge trajectory. This amplitude, determined by a pole at a given  complex $s$ and real $l$, can be extended to any value in the complex $s$-plane. In particular we can compare this  Regge amplitude on the real axis  with the partial waves of \cite{GarciaMartin:2011cn}.  The two amplitudes do not have to overlap on the real axis since they 
  are only constrained to agree at the resonance pole.

\begin{figure}
\includegraphics[scale=0.70,angle=-90]{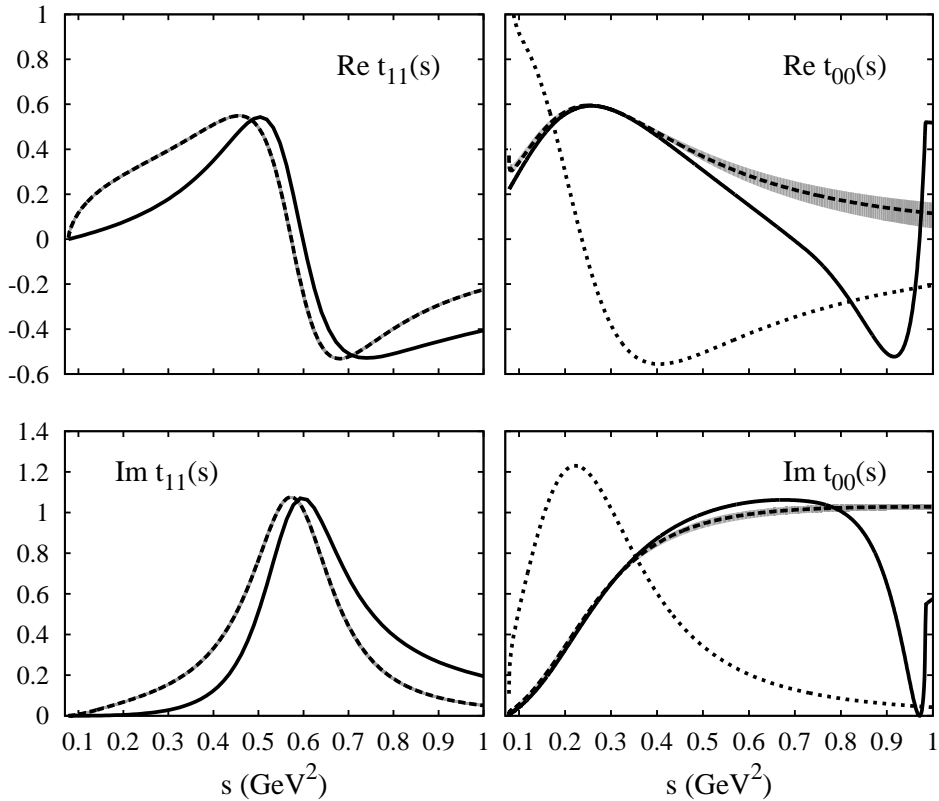}
 \caption{\rm \label{fig:ampl} 
  Partial waves $t_{lI}$  
  ($I$ being the isospin) with the $l=1$ wave shown in the left and $l=0$ in the right panels, respectively. 
  Solid lines represent the amplitudes from \protect{\cite{GarciaMartin:2011cn}}. The resonance poles of these amplitudes 
    \protect{\cite{GarciaMartin:2011jx}} determine   
   the  constrained Regge-pole amplitudes shown with dashed curves. The estimated systematic uncertainties are shown as gray bands (almost indistinguishable from the dashed line in the case of $t_{11}$.)
 In the right panels, the dotted lines represent the constrained Regge-pole amplitude for the $S$-wave 
if the $\sigma$-pole is fitted by  imposing a linear trajectory with $\alpha'\simeq 1\,$GeV$^{-2}$.}
\end{figure}

The left panels of Fig.\ref{fig:ampl} show the real and imaginary parts of the $P$ wave. The solid curves give the Constrained Data Fits 
of \cite{GarciaMartin:2011cn} whose $\rho(700)$ pole position, $\sqrt{s_\rho}=763.7^{+1.7}_{-1.5}-i73.2^{+1.0}_{-1.1}$ MeV and residue, $\vert g_\rho\vert=6.01^{+0.04}_{-0.07}$ \cite{GarciaMartin:2011jx} were used as input. As discussed above, these are then fit with Eq.~\eqref{Reggeliket} for $l=1$, where $\alpha(s)$ and $\beta(s)$ satisfy the coupled Eqs.\eqref{iteration1} and \eqref{iteration2}. 
The   output   values for the fitted pole are: $\sqrt{s_\rho}=762.7-i73.5$ MeV and $\vert g_\rho \vert=5.99$.  The resulting real and imaginary parts of this Regge-pole amplitude on the real axis 
are shown as dashed lines in the left column of Fig.\ref{fig:ampl}. The gray bands cover the uncertainties 
due to the errors in the determination of the pole positions and residues from 
 the dispersive analysis of data in \cite{GarciaMartin:2011cn}.
In the resonant region there is fair agreement between our resulting amplitude and that from \cite{GarciaMartin:2011cn}. 
The $\rho$ peak is clearly identified in the imaginary part, and, as expected, the agreement deteriorates 
as we approach threshold or the inelastic region, 
where the pole is less dominant. 

The right panels of Fig.\ref{fig:ampl} display the $S$ wave from \cite{GarciaMartin:2011cn} (solid curves), with a pole \cite{GarciaMartin:2011jx} at 
$\sqrt{s_\sigma}=457^{+14}_{-13}-i279^{+11}_{-7}\,$MeV 
and residue $\vert g_\sigma \vert=3.59^{+0.11}_{-0.13}$ GeV. Dashed lines correspond to the Regge-pole partial wave, whose pole is at $\sqrt{s_\sigma}=461-i281$ MeV and $\vert g_\sigma\vert=3.51$ GeV. 
It is well known that the $f_0(500)$  does not conform to a
 Breit-Wigner shape but still dominates the partial wave from threshold up to almost 1 GeV, where it strongly interferes with the very narrow $f_0(980)$. We find a remarkably good agreement between our input and output amplitudes from threshold up to $0.5$ GeV$^2$,
where the agreement starts to deteriorate.

\begin{figure}
\includegraphics[scale=0.70,angle=-90]{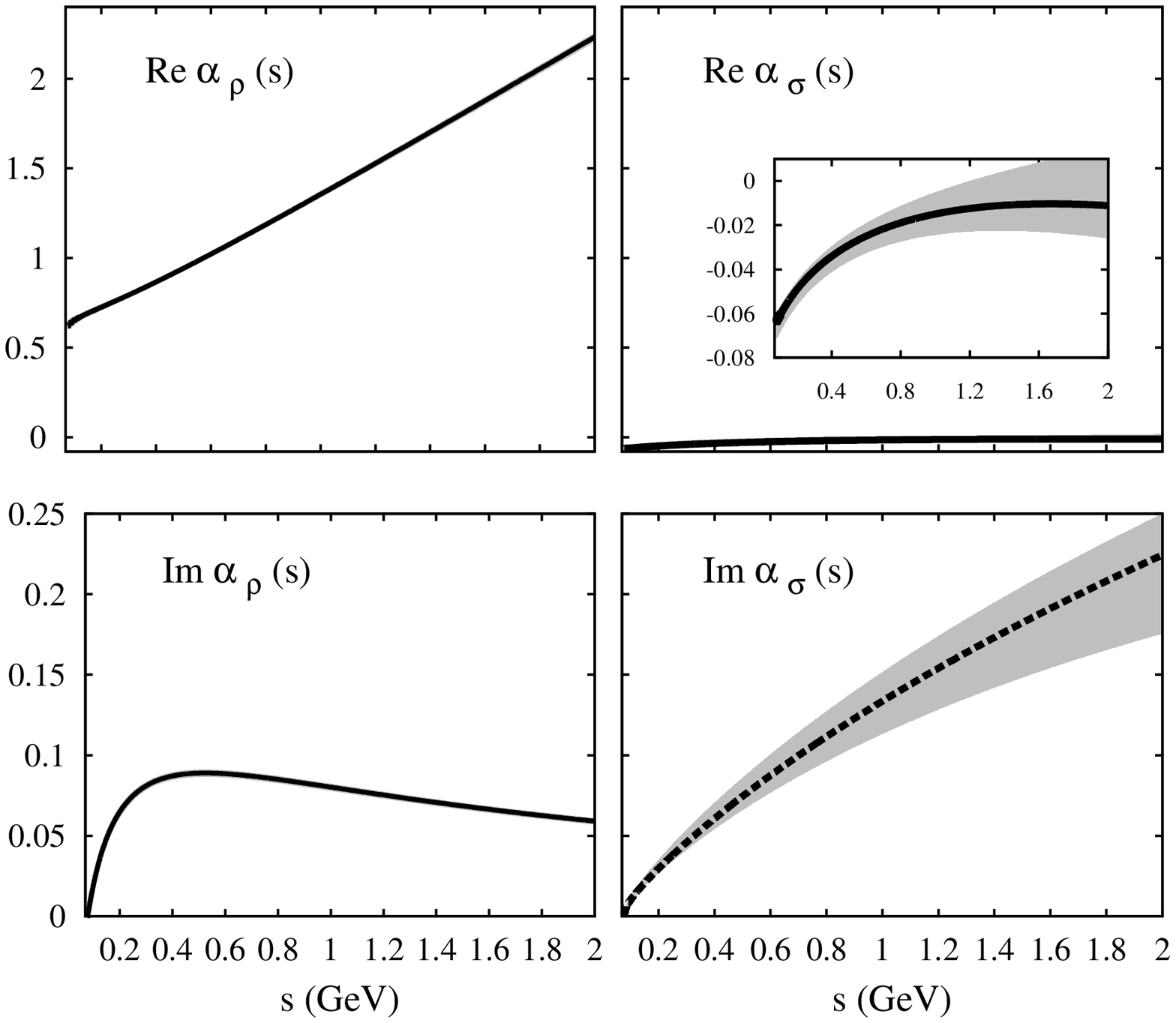}
 \caption{\rm \label{fig:trajectories1} 
   Real and imaginary parts of the resonance Regge trajectory obtained
from the resonance pole parameters as explained in the text. The $\rho(770)$ trajectory shown in the left panel  
 is almost real and linearly rising. In the right panel, we observe that
  real and imaginary parts of the $f_0(500)$  trajectory  are comparable with the former having a much smaller slope that that of 
 the $\rho$. }
\end{figure}

Since our constrained Regge amplitudes provide a fair representation of the resonance region,
we show in  Fig.\ref{fig:trajectories1} the resulting Regge trajectories. We see that 
the imaginary part of $\alpha_\rho(s)$ is much smaller than the real part. In addition, the latter grows linearly with $s$, with intercept 
$\alpha_\rho(0)=0.520\pm0.002$. 
Note that the error band we provide is only due to the uncertainty in the input pole parameters from ref.~\cite{GarciaMartin:2011jx}. 
  
This value for the  slope is consistent with that obtained from the
extensive study \cite{Anisovich:2000kxa} of $(J,M^2)$ resonance trajectories. 
It can also be compared  with $\alpha_\rho(0)=0.52\pm0.02$  from fits 
to total cross sections for $NN,\pi N$ and $\pi\pi$~\cite{Pelaez:2003ky},
or to the value of  $\alpha_\rho(0)=1-\eta_2=0.450\pm0.005$  
\cite{PDG}, which includes an even larger number of channels.    
Moreover, the resulting slope $\alpha'_\rho=0.902\pm0.004\,$GeV$^{-2}$ 
is consistent with fits 
of a linear trajectory to the $\rho(770)$, $\rho_3(1690)$ and $\rho_5(2350)$ mesons 
performed in \cite{Anisovich:2000kxa} and more recently in \cite{Masjuan:2012gc} that 
 yield $\alpha'_\rho\simeq 0.83\,$GeV$^{-2}$ and $\alpha'_\rho\simeq 0.87\pm0.06$GeV$^{-2}$, respectively. A 
 value of $\alpha'_\rho=0.9\,$GeV$^{-2}$ was used in \cite{Pelaez:2003ky}.
  
Taking into account our approximations,  and that our error bands only reflect
the uncertainty in the input pole parameters, our results 
 are in remarkable agreement with trajectories from the literature and provide a benchmark of the validity of our approach.  

 Regarding the $f_0(500)$ trajectory shown in the right panel of Fig.\ref{fig:trajectories1}, 
we see that it is evidently nonlinear. We obtain 
 $$\alpha_\sigma(0)=-0.090^{+0.004}_{-0.012},\quad \alpha'_\sigma\simeq 0.002^{+0.050}_{-0.001} \mbox{ GeV}^{-2},$$
where, once again, the error bands are due to the uncertainty in the input  pole parameters from \cite{GarciaMartin:2011jx}.
The slope is about two orders of magnitude
 smaller that  of the $\rho$ (and other trajectories typical to quark-antiquark  resonances, {\it e.g.}\ $a_2$, $f_2$, $\pi_2$). 
 This provides strong support for a non-ordinary nature of the $\sigma$ meson. Furthermore the  
 growth of $\alpha_\sigma(s)$ is so slow that it excludes the possibility 
 that any of the known isoscalar resonances $f_2, f_4,...$ lie on the $\sigma$ meson  trajectory.
Our result therefore explains why the $f_0(500)$ does not fit well in the 
 usual hadron classification into linear trajectories with a slope
 of typical hadronic size.

\begin{figure}
\includegraphics[scale=0.70,angle=-90]{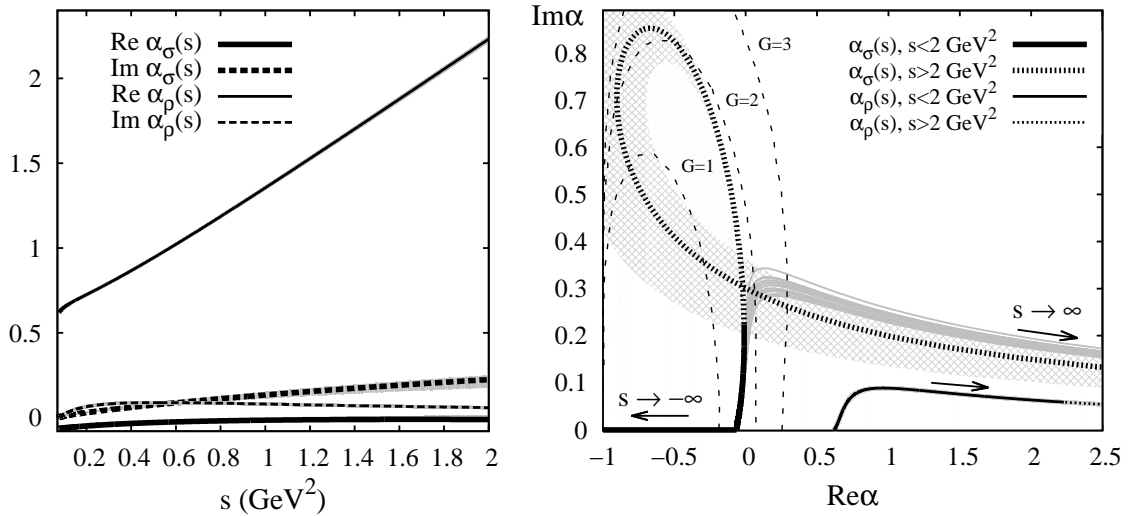}
 \caption{\rm \label{fig:trajectories} 
  (Left) $\alpha_\rho(s)$ and $\alpha_\sigma(s)$ Regge trajectories, 
from our constrained Regge-pole amplitudes. 
The $\rho$ trajectory is almost real and linear, whereas both the 
real and imaginary parts of $\alpha_\sigma(s)$ 
are very small, and not evidently linear. For the error bands we use the same convention as in Fig.\ref{fig:ampl}. In most cases the bands are so thin that they are barely distinguishable from the central lines. (Right) $\alpha_\sigma(s)$ and $\alpha_\rho(s)$
in the complex plane. Beyond $s=2\,$GeV$^2$ extrapolations of our results are plotted as dotted lines. 
Asymptotically, i.e., $s>200\,$GeV$^2$, the small real part, proportional to $\alpha' s$ takes over the dispersive contribution and 
 ${\rm Re}\,\alpha_\sigma(s)$ starts growing, similar to ${\rm Re}\,\alpha_\rho(s)$. 
 Within the input pole parameter errorbands, in the case of the $\sigma$, we find two types of solutions for the trajectory. One set 
  has a loop in the $\mbox{Im}\alpha - \mbox{Re}\alpha$ plane. The other, having slightly higher $\alpha'$ does not form a loop. 
The pattern-filled band encloses the first type of solutions,  whereas the gray lines correspond to the other set. 
At low and intermediate energies, both are similar to the trajectories of the Yukawa potential $V(r)={\rm G} \exp(-r/a) /(m_\pi a r)$, shown here for three different  values of  G  \cite{Lovelace}.  }
\end{figure}

To show the difference between the $\rho(770)$ and $f_0(500)$ trajectories,  in the left panel of Fig.\ref{fig:trajectories} 
we plot both the real and imaginary parts of the two trajectories using the same scale. Not only is the difference clearly evident  
between the shape and magnitude of the real part of the $\rho(770)$ trajectory
and that of the $f_0(500)$, but also the fact that the real and imaginary parts of the $f_0(500)$ trajectory
are comparable. 

Furthermore, in Fig.\ref{fig:trajectories} we show the striking similarities
 between the $f_0(500)$ trajectory and those of Yukawa potentials
in non-relativistic scattering, not only at low energies below $s=2\,$GeV$^2$, represented by the thick continuous line,
 but also when extrapolated  beyond that energy, which we show as a thick dashed-dotted line that describes a backward loop in the complex plane before moving to infinity.     Of course, our results are most reliable at low energies and the extrapolation should be interpreted cautiously. Nevertheless, our results suggest that the $f_0(500)$ looks more like a low-energy resonance of a short range potential,  {\it e.g.}\ between pions,  than a bound state of a long range confining force between a quark and an antiquark.

Concerning the uncertainties in the input parameters \cite{GarciaMartin:2011jx}, we observe that from threshold energies up to $s=2\,$GeV$^2$, i.e., the grey band around the thick continuous line, all trajectories bear a close similarity to Regge trajectories
of Yukawa potentials as it happens for the central curve.  Of course, when extrapolating our results to  higher energies, the uncertainty band becomes larger. Most of the trajectories we find within the uncertainties still describe a loop in the  (${\rm Re }\,\alpha, {\rm Im }\,\alpha$) plane, but a few of them describe a trajectory where the loop has collapsed (these are represented by a somewhat darker gray band). For the latter the $\alpha' s$ term is somewhat stronger and it prevents formation of a loop. 
  Below $s=2\,$GeV$^2$, all trajectories follow a qualitative behavior similar to that of a Yukawa potential and even
when extrapolated to higher energies they do {\it not} follow an ordinary almost-real linear Regge trajectory
with a slope of order $1\,$GeV$^{-2}$.

 One could also wonder if the weak $f_0(500)$ trajectory
 is affected by other uncertainties, hidden in the neglected background amplitude ({\it cf.}\ Eq.~(\ref{Reggeliket})). 
 If  we try to  fit the pole in~\cite{GarciaMartin:2011cn} 
 by fixing $\alpha'$ to a more natural value, say the one 
 for the $\rho(770)$, we obtain a $\chi^2$ which is two orders of magnitude larger, with the pole at
$\sqrt{s_\sigma}= 487 -i 199$ MeV and a much larger coupling 
$\vert g_\sigma\vert=4.09\,$GeV. Even worse, as seen by the dotted curve in the right panel of Fig.~\ref{fig:ampl}, on the real axis  
  the real and imaginary parts of the resulting Regge-pole amplitude 
 are qualitatively different from the expected behavior for the $S$-wave. 
  We note that this exercise is also highly relevant because it illustrates that the large resonance width is not responsible for the fact that the $f_0(500)$
 does not follow an ordinary Regge trajectory.

\begin{figure}
\includegraphics[scale=0.7,angle=-90]{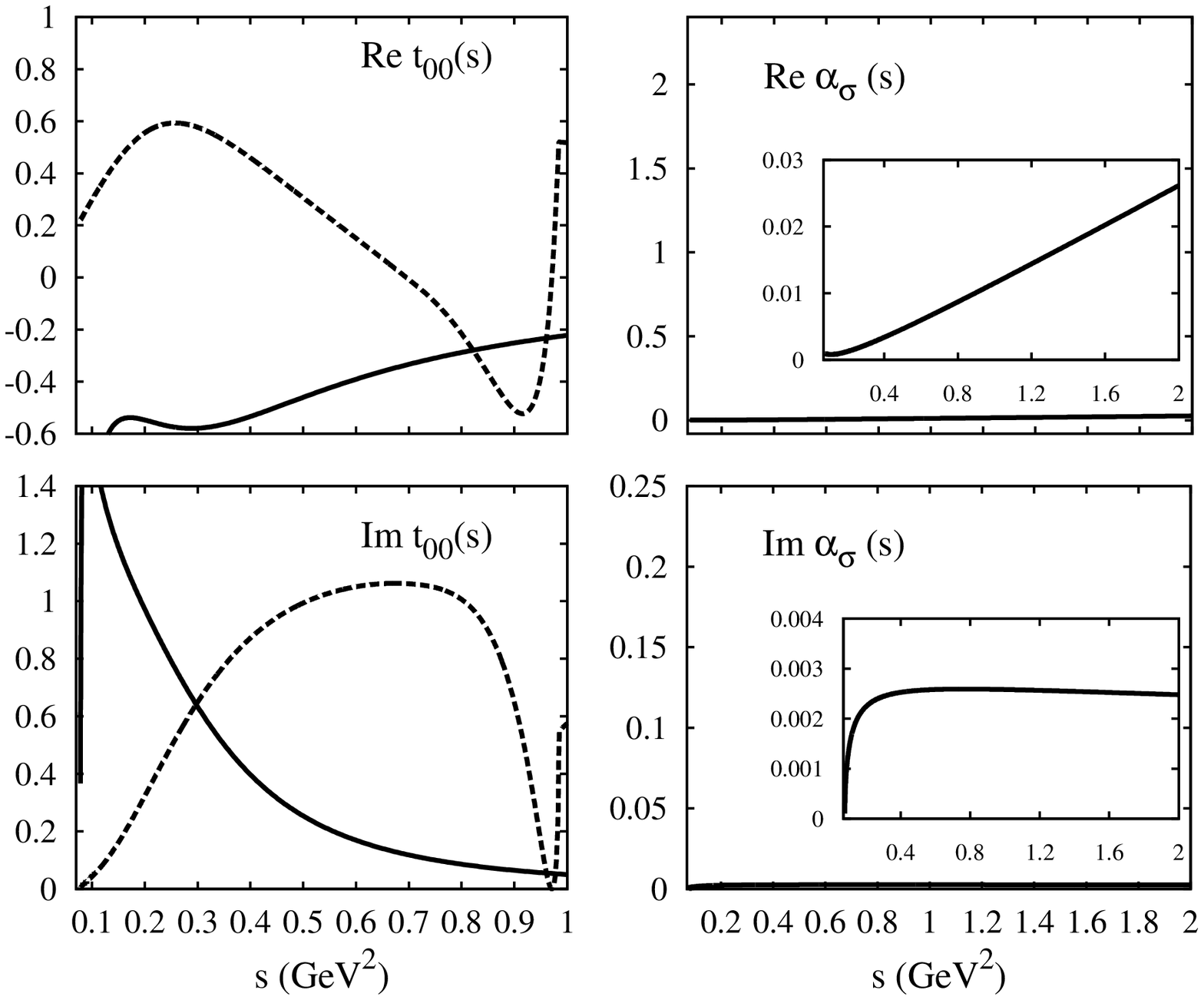}
 \caption{\rm \label{fig:sig-wo-adler} 
 Left panel:  Dashed lines represent real (top) and imaginary (bottom) parts of the dispersive data analysis in \protect{\cite{GarciaMartin:2011cn}} which provide the $\rho$ and $f_0(500)$ poles \protect{\cite{GarciaMartin:2011jx}}. These poles have been fitted here with the coupled dispersion relations of Eqs.~\eqref{iteration1} and \eqref{iteration2}, in which the Adler zero has not been imposed. The resulting real and imaginary parts of this Regge-pole amplitude are shown as black lines. Although the pole parameters are acceptably fitted, the amplitude in the real axis does not agree at all with that of the dispersive data analysis. Right panel: real (top) and imaginary (bottom) parts of the corresponding Regge trajectory. 
}
\end{figure}

Finally we show the results obtained for the $f_0(500)$ trajectory when using the unmodified dispersion relation of Eq.~\eqref{iteration2} instead of the one in which the Adler zero has been fixed. In this case the pole position and coupling can be reproduced fairly well: $\sqrt{s_\sigma}=476-i280$ MeV, $\vert g_\sigma\vert=3.20$ GeV (although with a $\chi^2$ still 20 times larger than that of the fit with the modified dispersion relation). However, the corresponding amplitude on the real energy axis turns out to be completely different from that of  the dispersive analysis, and consequently does not agree with the experimental data. This is apparent in the left panel of Fig.~\ref{fig:sig-wo-adler}. In addition, in the right panel of Fig.~\ref{fig:sig-wo-adler} we show the real and imaginary parts of its Regge trajectory. The results are quite similar to the case when we do not factor out the Adler zero. Actually,  both are once again completely flat compared to the real part of the $\rho$ trajectory, (to ease the comparison, the same scale as that of Fig.~\ref{fig:trajectories} has been used for the real part). The parameters of this trajectory are: $\alpha_0=-0.002$ and $\alpha'=0.015 \mbox{ GeV}^{-2}$, once again very different from the corresponding values for ordinary meson trajectories. It seems that the Adler zero is quite important to obtain a reasonable description of data, but
is not responsible for the non-ordinary behavior of the $f_0(500)$, which seems to be mainly determined by values of the pole parameters, i.e., the mass, width and residue.

\section{Summary and Discussion}

To summarize, we have shown how to obtain the Regge trajectory 
of a light resonance from its associated pole, when that dominates the elastic scattering of two 
hadrons. The method is based on general analytic properties and yields a set
of integral relations for the Regge trajectory and its residue.
These are solved iteratively while fitting just the pole position and coupling of the given resonance. The method works fairly well for the $\rho(770)$,
which dominates elastic $\pi\pi$ vector-isovector scattering.
The resulting trajectory is almost real and nearly linear. Given our approximations, the intercept and slope come remarkably close to values in the literature, obtained from fits to high energy scattering 
or to linear trajectories including the $\rho(770)$, $\rho_3(1690)$ and $\rho_5(2350)$. 
Our method thus identifies the $\rho(770)$ and its trajectory partners as ordinary mesons.
It is worth noting that, since higher resonances are not included in the input, our method
``predicts'' such a tower of resonances, from just the pole and residue of the $\rho(770)$.
Note that the method does not build in a nearly real and linear behavior. 

The main objective of this paper, however, was to estimate the Regge trajectory 
and residue of the $\sigma$ or $f_0(500)$ scalar meson, whose pole
position has been accurately determined by several groups 
using model independent dispersive techniques.
Our estimate is relevant because the $\sigma$ has been long considered a 
non-ordinary meson and is often excluded from
linear Regge fits with slopes of typical hadronic size, $\simeq 1\,$GeV$^{-2}$.

For the scalar case our method is modified to include the Adler zero required by chiral symmetry.
We fit the pole and coupling obtained from dispersive studies of
$\pi\pi$ scattering and obtain the Regge residue and trajectories. 
 The resulting trajectory is more than one order of magnitude weaker than 
that of the $\rho$ or any ordinary trajectory, and at low energies bears striking similarities
with the trajectories of Yukawa potentials.
The resulting scale of tens of MeV or at most hundreds, for the slope, is more typical of meson physics than of quark-antiquark interactions.  
The $\sigma$ Regge trajectory is so flat that any trajectory partners  
would have to be extremely massive.
To test the robustness of this observation we have checked that our results are very stable 
within the uncertainties of the pole parameters that we used as input. In addition we have
tried to impose a typical size linear trajectory on the $\sigma$, but that deteriorates
the fit to the $\sigma$ pole and particularly to the coupling,
so the resulting amplitude in the physical region is qualitatively very different from the observations. 
Therefore, the smallness of our estimate of the $\sigma$ trajectory is robust and explains why it does not fit well in the usual Regge classification and strongly supports a non-ordinary nature of
the lightest scalar meson.

  Our method can be  applicable to other resonances that dominate  elastic scattering and generalization to inelastic channels is also straightforward.    Hence we plan on studying the $K^*(892)$ vector and the $K(800)$ (or $\kappa$) scalar resonance, and possibly other meson-nucleon resonances. Furthermore we also plan on studying the $N_c$ or quark mass dependence of the Regge trajectories   and  to explore hadronic resonance models that could explain this non-ordinary behavior (e.g., tetraquarks, hadron molecules, etc).  We expect that this method should provide further understanding of the most controversial
states in the hadron spectrum.
   
\section*{Acknowledgments} JRP and JN are supported by the Spanish project FPA2011-27853-C02-02. JN acknowledges funding by the Deutscher Akademischer Austauschdienst (DAAD), the Fundaci\'on Ram\'on Areces and the hospitality of Bonn and Indiana Universities. 
APS\ is supported in part by the U.S.\ Department of Energy under Grant DE-FG0287ER40365. JTL is supported by the U.S. National Science Foundation under grant 
PHY-1205019.


\vspace*{-.2cm}
\begin{thebibliography}{99}
\vspace*{-.2cm}

\bibitem{Jaffe:1976ig}
  R.~L.~Jaffe,
  Phys.\ Rev.\  D {\bf 15}, 267 (1977);
  Prog.\ Theor.\ Phys.\ Suppl.\  {\bf 168}, 127 (2007).
%
%
  J.~D.~Weinstein and N.~Isgur,
  Phys.\ Rev.\ Lett.\  {\bf 48}, 659 (1982).
  U.~G.~Meissner,
  Comments Nucl.\ Part.\ Phys.\  {\bf 20}, 119 (1991).
  E.~van Beveren,{\it et al.} 
  Z.\ Phys.\ C {\bf 30}, 615 (1986).
%
  D.~Black,
  A.~H.~Fariborz, F.~Sannino and J.~Schechter,
  Phys.\ Rev.\ D {\bf 59}, 074026 (1999).
  P.~Minkowski and W.~Ochs,
  Eur.\ Phys.\ J.\  C {\bf 9}, 283 (1999).
%
%
  J.~A.~Oller and E.~Oset,
  Nucl.\ Phys.\  A {\bf 620}, 438 (1997)
  [Erratum-ibid.\  A {\bf 652}, 407 (1999)].
%
%
E.~van~Beveren and G.~Rupp,
  Eur.\ Phys.\ J.\  C {\bf 22}, 493 (2001).
%
  F.~E.~Close and N.~A.~Tornqvist,
  J.\ Phys.\ G {\bf 28}, R249 (2002)
%
  J.~R.~Pelaez,
  Phys.\ Rev.\ Lett.\  {\bf 92}, 102001 (2004).
%
  J.~Vijande, 
  A.~Valcarce, F.~Fernandez and B.~Silvestre-Brac,
  Phys.\ Rev.\  D {\bf 72}, 034025 (2005).
%
  F.~Giacosa,
  Phys.\ Rev.\ D {\bf 74}, 014028 (2006).
%
  L.~Maiani, A.~D.~Polosa and V.~Riquer,
  Phys.\ Lett.\ B {\bf 651}, 129 (2007).
%
  G.~'t Hooft, {\it et al.} 
  Phys.\ Lett.\ B {\bf 662}, 424 (2008).
%
  T.~Hyodo, D.~Jido and T.~Kunihiro,
  Nucl.\ Phys.\  A {\bf 848}, 341 (2010).
%


\bibitem{Pelaez:2006nj} 
  J.~R.~Pelaez and G.~Rios,
  Phys.\ Rev.\ Lett.\  {\bf 97}, 242002 (2006).
J.~R.~Pelaez, 
M.~R.~Pennington,  J.~Ruiz de Elvira, and D.~J.~Wilson,
  Phys.\ Rev.\ D {\bf 84}, 096006 (2011).
  %


\bibitem{Anisovich:2000kxa} 
  A.~V.~Anisovich, V.~V.~Anisovich and A.~V.~Sarantsev,
  Phys.\ Rev.\ D {\bf 62}, 051502 (2000).

\bibitem{Masjuan:2012gc} 
  P.~Masjuan, E.~Ruiz Arriola and W.~Broniowski,
  Phys.\ Rev.\ D {\bf 85}, 094006 (2012).



\bibitem{Caprini:2005zr}
  I.~Caprini, G.~Colangelo and H.~Leutwyler,
  Phys.\ Rev.\ Lett.\  {\bf 96},  132001 (2006).

\bibitem{GarciaMartin:2011jx} 
  R.~Garcia-Martin, 
  R.~Kaminski, J.~R.~Pelaez and J.~Ruiz de Elvira,
  Phys.\ Rev.\ Lett.\  {\bf 107}, 072001 (2011).



\bibitem{PDG}
J. Beringer et al. (Particle Data Group), Phys. Rev. {\bf D86}, 010001 (2012). 
  J.~R.~Pelaez,
  PoS ConfinementX {\bf }, 019 (2012)
  [arXiv:1301.4431 [hep-ph]].


\bibitem{Chu:1969ga} 
G. Epstein and P. Kaus, Phys. Rev. {\bf 166}, 1633 (1968);
  S.~-Y.~Chu, 
  G.~Epstein, P.~Kaus, R.~C.~Slansky and F.~Zachariasen,
  Phys.\ Rev.\  {\bf 175}, 2098 (1968).
  

\bibitem{Collins-PLB} P.D.B.~Collins, R.C.~Johnson and E.J.~Squires, Phys.\ Lett. B{\bf 26}, 223 (1968). 

\bibitem{GarciaMartin:2011cn} 
  R.~Garcia-Martin, 
  R.~Kaminski, J.~R.~Pelaez, J.~Ruiz de Elvira and F.~J.~Yndurain,
  Phys.\ Rev.\ D {\bf 83}, 074004 (2011).


\bibitem{Pelaez:2003ky} 
  J.~R.~Pelaez and F.~J.~Yndurain,
  Phys.\ Rev.\ D {\bf 69}, 114001 (2004).

\bibitem{chpt}
S. Weinberg, Physica {\bf A96}, 327 (1979).
J.~Gasser and H.~Leutwyler,
Annals Phys.\  {\bf 158}, 142 (1984).


\bibitem{Lovelace}  C. Lovelace and D. Masson, Nuovo Cimento {\bf 26} 472 (1962).
A. O. Barut and F. Calogero,  Phys.\ Rev.\ {\bf 128}, 1383 (1962). 
A. Ahamadzadeh, P.G. Burke and C. Tate,  Phys.\ Rev.\ {\bf 131} 1315 (1963). 

\end{thebibliography}
\end{document}